\documentclass[twocolumn,showpacs,preprintnumbers,amsmath,amssymb]{revtex4}
\pdfoutput=1
\usepackage{color,latexsym,epsfig,amssymb,amsfonts,amsmath,graphicx,bbm,enumerate,bm}
\usepackage{graphicx}

\newcommand{\ket}[1]{\left | \, #1 \right \rangle}

\newcommand{\mi}{\mathrm{i}}
\newcommand{\rmd}{\mathrm{d}}

\begin{document}

\preprint{APS/123-QED}

\title{Quantum memory in an optical lattice}% Force line breaks with \\

\author{J.~Nunn, U.~Dorner, P.~Michelberger, K.~Reim, K.~C.~Lee, N.~K.~Langford, I.~A.~Walmsley and D.~Jaksch}

\affiliation{Clarendon Laboratory, University of Oxford, Parks Road,
Oxford OX1 3PU, United Kingdom}

\date{\today}% It is always \today, today,

\begin{abstract}
Arrays of atoms trapped in optical lattices are appealing as storage media for photons, since motional dephasing of the atoms is eliminated. The regular lattice is also associated with band structure in the dispersion experienced by incident photons. Here we study the influence of this band structure on the efficiency of quantum memories based on electromagnetically induced transparency (EIT) and on Raman absorption. We observe a number of interesting effects, such as both reduced and superluminal group velocities, enhanced atom-photon coupling and anomalous transmission. These effects are ultimately deleterious to the memory efficiency, but they are easily avoided by tuning the optical fields away from the band edges.
\end{abstract}

%Optical implementations of quantum information processing and transfer
%Quantum description of interaction of light and matter; related experiments}
%Quantum optics

\maketitle
\section{Introduction}
Quantum memories for photons based on coherent absorption in atomic ensembles promise to provide the light-matter interface required for scalable quantum optical networking \cite{Choi:2008xi,Hosseini:2009lq,Reim:2010kx,Novikova:2008ai,Phillips:2008ly}. Recently, arrays of atoms trapped in optical lattices \cite{Jaksch:1998zr} have been used as a storage medium \cite{Schnorrberger:2009ao,Dudin:2010fk}. The lattice eliminates decoherence via atomic diffusion and collisions, and coherence times of many seconds are feasible \cite{Lundblad:2010uq}. On the other hand, the periodic arrangement of atoms is expected to generate a photonic band structure for incident signals, with certain optical frequencies forbidden due to the destructive interference of scattered fields \cite{Petrosyan:2007vn,Yablonovitch:1987uq,Knight:1996kx}. At frequencies close to the edge of a forbidden band, the group velocity of an incident signal is reduced, and the signal therefore interacts with the atoms for longer \cite{Petrosyan:2007vn,Florescu:2005ys}. It is not clear what this implies for the efficiency of optical lattice quantum memories compared with their free-atom counterparts. In this paper we model quantum storage in a periodically structured ensemble in order to investigate this question. Our results indicate that while the interaction of the signal with the atoms grows stronger near a band edge, the memory efficiency is \emph{reduced}, because the coupling to the storage state of the memory is not enhanced. While the lattice band structure is not advantageous for quantum memory, simply tuning the optical fields away from the band edges allows for efficient storage, with all the benefits of reduced decoherence that accrue from the ability to fix the atoms in space.

\section{Model}
Optical lattices are capable of supporting highly entangled quantum states, ranging from atomic superfluids to Mott insulators \cite{Jaksch:1998zr}. However, a semiclassical treatment in which the optical fields and the atomic positions remain un-quantized is sufficient to analyze the efficiency of quantum storage, which quantity involves only normally ordered products of field amplitudes. The following analysis therefore does not depend on the spatial correlations within the lattice, and so it also applies to any ensemble memory in which the atomic density varies periodically in space on the scale of an optical wavelength, such as might be produced by doping a photonic crystal structure \cite{Florescu:2005ys}.

\begin{figure}[h]
\begin{center}
\includegraphics[width=\columnwidth]{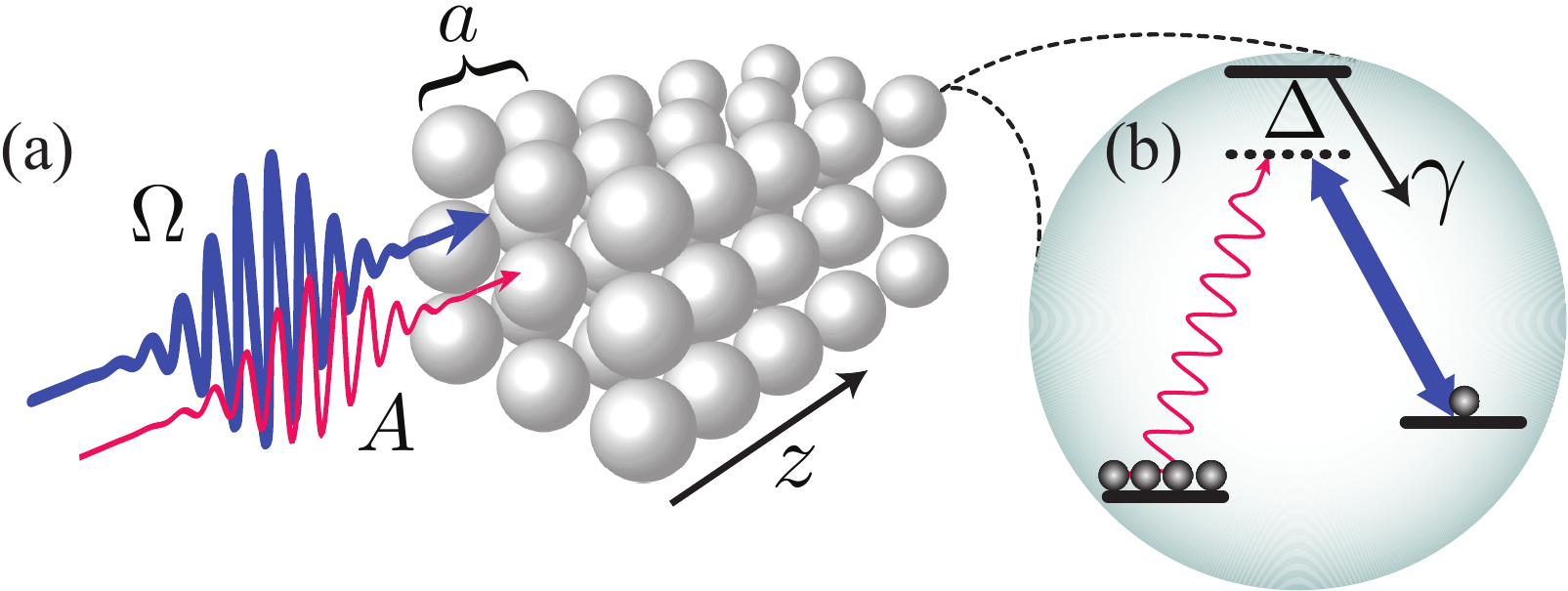}
\caption{(Color online) Quantum memory in an optical lattice. (a) A signal field $A$ is directed into an ensemble of atoms trapped in a regular array, along with a bright control field $\Omega$. (b) Each atom has a $\Lambda$-level structure. The control and signal fields are tuned into two-photon resonance, precipitating the transfer of atoms from the the ground state $\ket{1}$ to the long-lived storage state $\ket{3}$. In an EIT memory, the common detuning $\Delta$ of the signal and control from the excited state $\ket{2}$ is zero; in a Raman memory, $\Delta\gg \gamma$, where $\gamma$ is the homogeneous linewidth of the $\ket{1}\leftrightarrow\ket{2}$ transition.}
\label{fig:lattice3}
\end{center}
\end{figure}

We consider quantum memories based on EIT \cite{Fleischhauer:2000vn,Novikova:2008ai} and Raman scattering \cite{Kozhekin:2000bs,Reim:2010kx,Nunn:2007wj} in an ensemble of $\Lambda$-type atoms (see figure \ref{fig:lattice3}). In these memory protocols, the signal photon to be stored is absorbed on the $\ket{1}\leftrightarrow\ket{2}$ transition, and mapped into the storage state $\ket{3}$ by an intense control field. In a Raman memory the signal and control are tuned far from resonance by a common detuning $\Delta$; the fields are tuned into resonance, with $\Delta = 0$, in an EIT memory. A single theoretical model therefore suffices to describe both protocols \cite{Gorshkov:2007qm,Surmacz:2008vf}.
We consider propagation of the signal in one dimension, along the $z$-axis. The signal electric field $E_s$ couples to the atomic polarization $P_s$ via the wave equation
\begin{equation}
\label{Wave}
\left[\partial_z^2 -\frac{1}{c^2}\partial_t^2\right]E_s =-\mu_0\partial_t^2 P_s.
\end{equation}
Bragg scattering of the signal from the periodic array of atoms in the optical lattice introduces backward travelling components into the signal beam. We account for this by introducing a carrier wave for the signal $\phi$, which is not in general a plane wave. We introduce the slowly varying amplitude $A$ for the signal by the relation
\begin{equation}
\label{Es}
E_s(z,t) = \mi g_sA(z,t)\phi(z)e^{-\mi \omega_s t},
\end{equation}
where $g_s = \sqrt{\hbar \omega_s/2 \epsilon_0 \mathcal{A}c}$, with $\mathcal{A}$ the cross-sectional area of the signal field, and where $\omega_s$ is the signal carrier frequency. We describe the modulation of the atomic density due to the lattice with the periodic function $m(z) = m(z+a)$, where $a$ is the lattice constant, and where $\int_0^L m(z)\,\rmd z = L$, with $L$ the length of the ensemble, so that $m=1$ describes a uniform ensemble with no optical lattice. The slower spatial variation of the atomic polarization over the length of the ensemble is described by the amplitude $P$, which we define by the relation
\begin{equation}
\label{Ps}
P_s(z,t) = \frac{\sqrt{n}d_{12}}{\mathcal{A}}m(z)P(z,t)\phi(z)e^{-\mi \omega_s t},
\end{equation}
where we have factorized out both the lattice modulation, and the signal carrier wave $\phi$. Here $n$ is the constant average number density of atoms in the ensemble, and $d_{12}$ is the dipole matrix element for the $\ket{1}\leftrightarrow\ket{2}$ transition. The atomic dynamics in the presence of the signal and control fields are described by the Bloch equations, which can be written in the form \cite{Gorshkov:2007qm,Surmacz:2008vf}
\begin{eqnarray}
\label{P} \partial_t P &=& -\Gamma P + \mi \kappa A + \mi \Omega B,\\
\label{B} \partial_t B &=& \mi \Omega^* P,
\end{eqnarray}
where $\Gamma = \gamma-\mi \Delta$ is the complex detuning, with $\gamma$ the homogeneous linewidth of the excited state $\ket{2}$, and where the coupling constant is $\kappa = \sqrt{d\gamma/L}$, with $d=d_{12}^2\omega_snL/2 \epsilon_0 c\hbar \gamma$ the resonant optical depth of the ensemble. $B$ represents the amplitude of the long-lived $\ket{1}\leftrightarrow\ket{3}$ Raman coherence, or \emph{spin wave}, into which the signal field is mapped by the memory interaction. The slowly varying Rabi frequency $\Omega = \Omega(t-z/c)$ represents the temporal profile of the control pulse, which experiences no dispersion and travels at $c$, since it couples the states $\ket{2}$ and $\ket{3}$, whose populations remain negligible at all times.

When the spectral bandwidth $\delta$ of the control field is sufficiently narrow, $\delta \ll \Delta$, or $\delta \ll d\gamma$, the atomic polarization $P$ can be \emph{adiabatically eliminated} \cite{Raymer:1981zh,Duan:2002ul}. This is achieved by setting the time derivative on the left hand side of Eq.~(\ref{P}) to zero, and substituting the resulting solution for $P$ into Eqs.~(\ref{B}) and (\ref{Wave}), using Eq.~(\ref{Ps}). We make a slowly varying envelope approximation by neglecting the terms $\partial_z^2 A$, $\partial_t^2 A$ and both $\partial_t P$ and $\partial_t^2 P$ in Eq.~(\ref{Wave}). This yields the pair of coupled equations
\begin{eqnarray}
\nonumber 2\partial_z A \partial_z \phi + A\partial_z^2\phi & &\\
\label{Aphi} +k_s^2V A\phi+\frac{2\mi k_s}{c}\phi\partial_t A  &=& 2 k_sm\Omega\frac{\kappa}{\Gamma}B\phi,\\
\label{Bphi} \left(\partial_t + \frac{|\Omega|^2}{\Gamma}\right)B &=& \mi\Omega^*\frac{\kappa}{\Gamma}A,
\end{eqnarray}
where $V = V(z) = 1+\frac{2\mi d\gamma m(z)}{\Gamma L k_s}$, with $k_s = \omega_s/c$.
These equations simplify considerably if we require that the carrier wave $\phi$ should satisfy the Schr\"{o}dinger-like equation
\begin{equation}
\label{eig}
\left[\partial_z^2 + k_s^2V(z)\right]\phi(z) = 0.
\end{equation}
Here $V$ plays the role of a potential, which takes the form of a constant with a small modulation added to it. The periodicity of the potential ensures that the carrier wave can be expressed in the Bloch-Floquet form \cite{Kittel:1996ly}
\begin{equation}
\label{carrier}
\phi(z)=e^{\mi k z}u_{k\nu}(z),
\end{equation}
where $k$ is the \emph{crystal momentum} and where $u_{k\nu}(z) = u_{k\nu}(z+a)$ is a periodic \emph{Bloch function}. Substituting Eq.~(\ref{carrier}) into Eq.~(\ref{eig}), we find that the band index $\nu$ must be chosen along with the crystal momentum $k$ in order to satisfy the equation
\begin{equation}
\label{bloch}
\left[\partial_{z}^2 + 2\mi k \partial_{z} - k^2 + k_s^2V(z)\right]u_{k\nu}(z)=0.
\end{equation}
If it is not possible to find values of $k$ and $\nu$ that satisfy Eq.~(\ref{bloch}), the signal frequency $\omega_s$ is said to lie within a photonic band gap. We will be interested in frequencies that lie close to the edge of a band gap.

When Eq.~(\ref{bloch}) is satisfied, Eq.~(\ref{Aphi}) becomes
\begin{equation}
\label{Aphi2}
\partial_z A\partial_z\phi + \frac{\mi k_s}{c}\phi \partial_t A = k_s m\Omega \frac{\kappa}{\Gamma}B\phi.
\end{equation}
The potential $V$ contains the complex detuning, whose imaginary part describes absorption of the signal field by the atoms. In general, therefore, the crystal momentum $k$ has a small but non-zero imaginary part, and the carrier wave is damped to some degree. It is convenient, in analyzing the memory efficiency, to transfer this damping to the slowly varying signal amplitude $A$, which we do by making the transformations $A\longrightarrow Ae^{-\mathrm{Im}\left\{k\right\}z}$ and $\phi\longrightarrow \phi e^{\mathrm{Im}\left\{k\right\}z}=e^{\mi \mathrm{Re}\left\{k\right\}z}u_{k\nu}$.

A final simplification is achieved by `projecting' Eq.~(\ref{Aphi2}) onto the mode $\phi$ \cite{Florescu:2005ys}. This is accomplished by multiplying both sides of Eq.~(\ref{Aphi2}) by a \emph{conjugate} mode $\psi$, and integrating over a unit cell of the lattice, using the fact that the functions $A$, $B$ and $\Omega$ are effectively constant over this range. The conjugate mode $\psi$ is not equal to $\phi^*$, because Eq.~(\ref{eig}) is not generally a Hermitian eigenvalue equation. Re-writing Eq.~(\ref{eig}), we have that $\phi$ is the right eigenvector of the operator $M=-V^{-1}\partial_z^2$, with eigenvalue $k_s^2$. The corresponding left eigenvector of $M$ is then equal to $\psi$.

After these manipulations, we arrive at the following set of equations of motion for the quantum memory
\begin{eqnarray}
\label{motionA} \left[\partial_z + \mathrm{Im}\left\{k\right\} \right]A&=&-\left(c/v_\mathrm{g}\right)\alpha\frac{\mi  \kappa}{\Gamma}\Omega(\tau+\beta z)B,\\
\label{motion}\left[\partial_\tau +  \frac{|\Omega(\tau+\beta z)|^2}{\Gamma} \right]B &=& \frac{ \mi\kappa}{\Gamma}\Omega^*(\tau+\beta z)A,
\end{eqnarray}
where $\tau = t-z/v_\mathrm{g}$ denotes the time in a frame moving at the signal group velocity, given by \cite{Florescu:2005ys}
\begin{equation}
\label{group} v_\mathrm{g} = \frac{c}{k_s}\int_0^a \psi(z)\left[-\mi \partial_z \phi(z)\right] \,\rmd z,
\end{equation}
and where $\beta = 1/v_\mathrm{g}-1/c$ quantifies the rate at which the control walks off from the signal \cite{Wasilewski:2006th}. The \emph{overlap} $\alpha$ captures the degree to which the optical Bloch mode is matched to the lattice modulation
\begin{equation}
\label{overlap}
\alpha = \int_0^a \psi(z)\phi(z) m(z)\,\rmd z.
\end{equation}
Note that far from a band edge, the carrier wave $\phi$ takes the form of a plane wave, so that $\alpha=1$, $v_\mathrm{g}=c$ and $\beta=0$. The equations then reduce to the standard equations describing the Raman and EIT protocols in a disordered ensemble \cite{Gorshkov:2007qm,Nunn:2007wj}.
\section{Efficiency}
To analyze the memory efficiency, we need to solve the coupled system Eqs.~(\ref{motionA}), (\ref{motion}) and construct the Green's function $K$ such that \cite{Wasilewski:2006th}
\begin{equation}
\label{green}
B_\mathrm{out}(z) = \int_{-\infty}^\infty K(z,\tau)A_\mathrm{in}(\tau)\,\rmd \tau,
\end{equation}
where $B_\mathrm{out}(z) = B(z, \tau\rightarrow \infty)$ is the spin wave left in the atoms at the end of the storage interaction, and where $A_\mathrm{in}(\tau) = A(z=0,\tau)$ is the amplitude of the signal impinging on the entrance face of the ensemble. The storage efficiency is defined as the ratio of the number of final spin wave excitations to the number of incident photons, which is given by
\begin{equation}
\label{eff}
\eta = \frac{\int_0^L |B_\mathrm{out}(z)|^2\,\rmd z}{\int_{-\infty}^\infty |A_\mathrm{in}(\tau)|^2\,\rmd \tau}.
\end{equation}
The optimal storage efficiency $\eta_\mathrm{opt}$ can be found from the singular value decomposition \cite{Trefethen:1997sf} of the Green's function $K$ by squaring the largest singular value \cite{Nunn:2008xr}.

In general, the equations of motion cannot be solved analytically, because of the walk-off between the signal and the control field. However, the typical size of optical lattices ($\sim 1$ mm) is small compared with the longitudinal spatial extent $Tc$ of a typical photonic wavepacket, where $T$ is the signal pulse duration, for all but the shortest signal pulses. In this case, even very close to a band edge, we have $\beta L\ll T$, and we can safely drop $\beta$ from Eqs.~(\ref{motionA}), (\ref{motion}). The Green's function is then given by \cite{Nunn:2007wj,Kozhekin:2000bs,Gorshkov:2007qm}
\begin{equation}
\label{Green_give}
K(z,\tau) = \frac{\sqrt{\kappa}}{\Gamma} \Omega(\tau)e^{-\chi(z,\tau)}J_0\left(2\sqrt{\alpha\kappa z\omega(\tau)c/v_\mathrm{g}}/\Gamma\right),
\end{equation}
where $\chi(z,\tau)=\mathrm{Im}\left\{k\right\}z+\omega(\tau)/\Gamma$, and where we have defined the integrated Rabi frequency $\omega(\tau) = \int_{-\infty}^\tau \left|\Omega(\tau')\right|^2\,\rmd\tau'$. Here $J_0$ denotes the zero'th order ordinary Bessel function of the first kind. A coordinate transformation from $\tau$ to $\omega = \omega(\tau)$ reveals that the singular values of this kernel do not depend on the temporal profile $\Omega$; they depend only upon the energy in the control pulse, represented by the limit $\omega(\tau\rightarrow \infty)$. Therefore the optimal storage efficiency for a lattice memory depends on the control pulse energy, but not on its shape.

To understand the effect of the photonic band structure induced by the optical lattice, we study the variation of the optimal efficiency $\eta_\mathrm{opt}$ as we vary the lattice constant $a$, with all other quantities fixed. In the calculations, we assume atomic parameters similar to those for cesium or rubidium, so we set $\lambda_s = 2\pi/k_s = 800$ nm, and $\tau_2 = 1/\gamma = 30$ ns. We assume that the atoms are trapped in an optical lattice of length $L=1$ mm, and of width sufficient to cover the full beam waist of the signal field. We assume a Gaussian profile $m(z) \propto e^{-(z/w)^2}$ for the density modulation due to the lattice potential, with width $w = a/10$. For concreteness, a Gaussian profile $\Omega(\tau)  = \Omega_0e^{-(\tau/T)^2}$ is assumed for the control pulse, with a peak Rabi frequency $\Omega_0 = 5.5/T$.

\begin{figure}[h]
\begin{center}
\includegraphics[width=7cm]{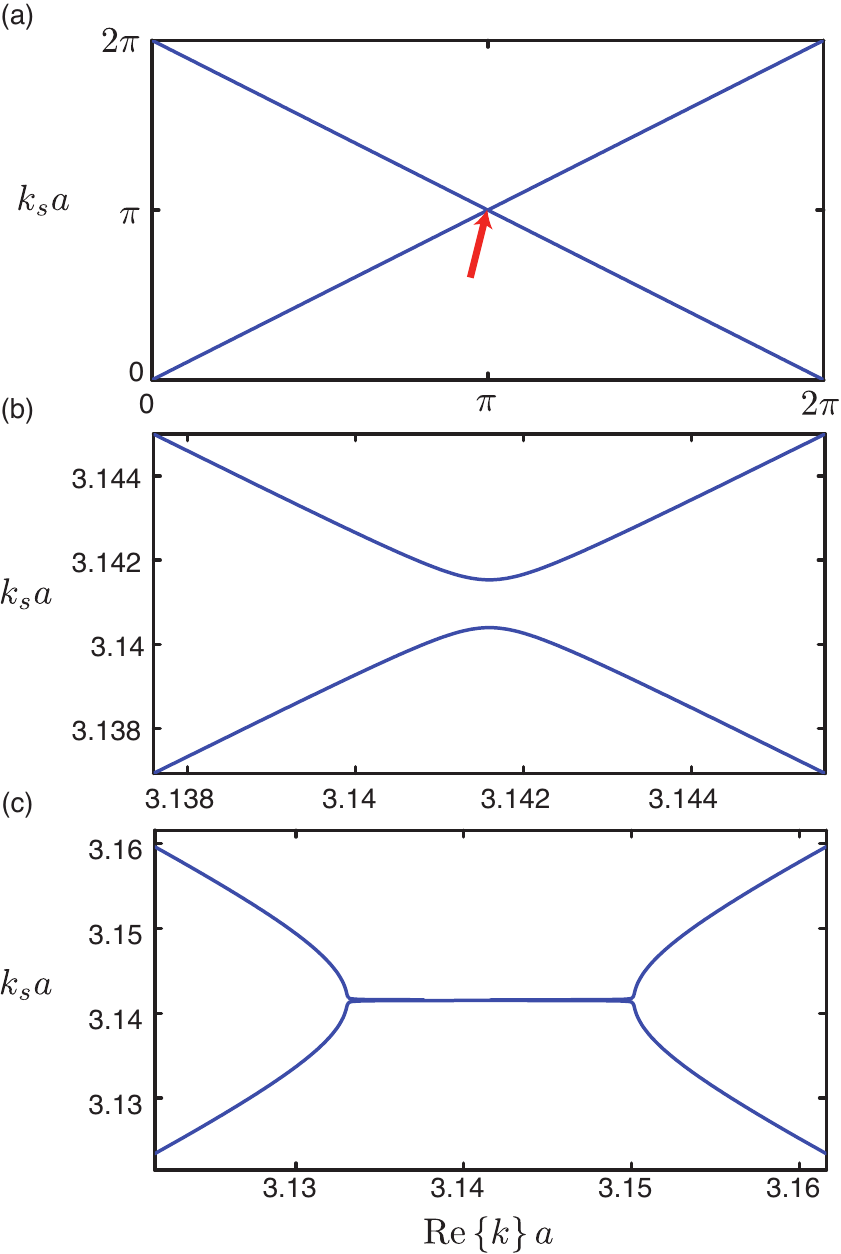}
\caption{(Color online) Dispersion in an optical lattice. (a) The dispersion relation for the signal field: that is, the dependence of the Bloch wave eigenvalue $k_s$ upon the real part of the crystal momentum $k$. The dispersion relation is multi-valued; for a given crystal momentum, the eigenvalues are enumerated by the band index $\nu$. The bands with $\nu=1$ and $\nu=2$ are shown here. The band gaps appear at the position indicated by the arrow, which marks the edge of the first Brillouin zone. (b) Close-up of the first Brillouin zone edge for the case of a Raman memory. The band gap has the appearance of a `standard' anti-crossing. (c) Corresponding close-up for an EIT memory. The band structure here is more complex, because the modulation on the potential $V(z)$ is imaginary. The gradient of the $\nu=1$ band increases at first, before sharply approaching zero. The bandgap itself is too small to see in the plot.}
\label{fig:dispersion}
\end{center}
\end{figure}

We consider two situations. First, a broadband off-resonant Raman protocol, with $d=300$, $T=3$ ns and $\Delta = 15/T$, and second, a narrowband resonant EIT protocol with $d=30$, $T=30$ ns and $\Delta = 0$ \cite{Surmacz:2008vf}. Note that the adiabatic condition $Td\gamma\gg 1$ is well-satisfied in both cases \cite{Gorshkov:2007qm}. Both types of memory exhibit band structure (Figure~\ref{fig:dispersion}), but the dispersive properties of Raman and EIT memories are quite different.

\begin{figure}[h]
\begin{center}
\includegraphics[width=7cm]{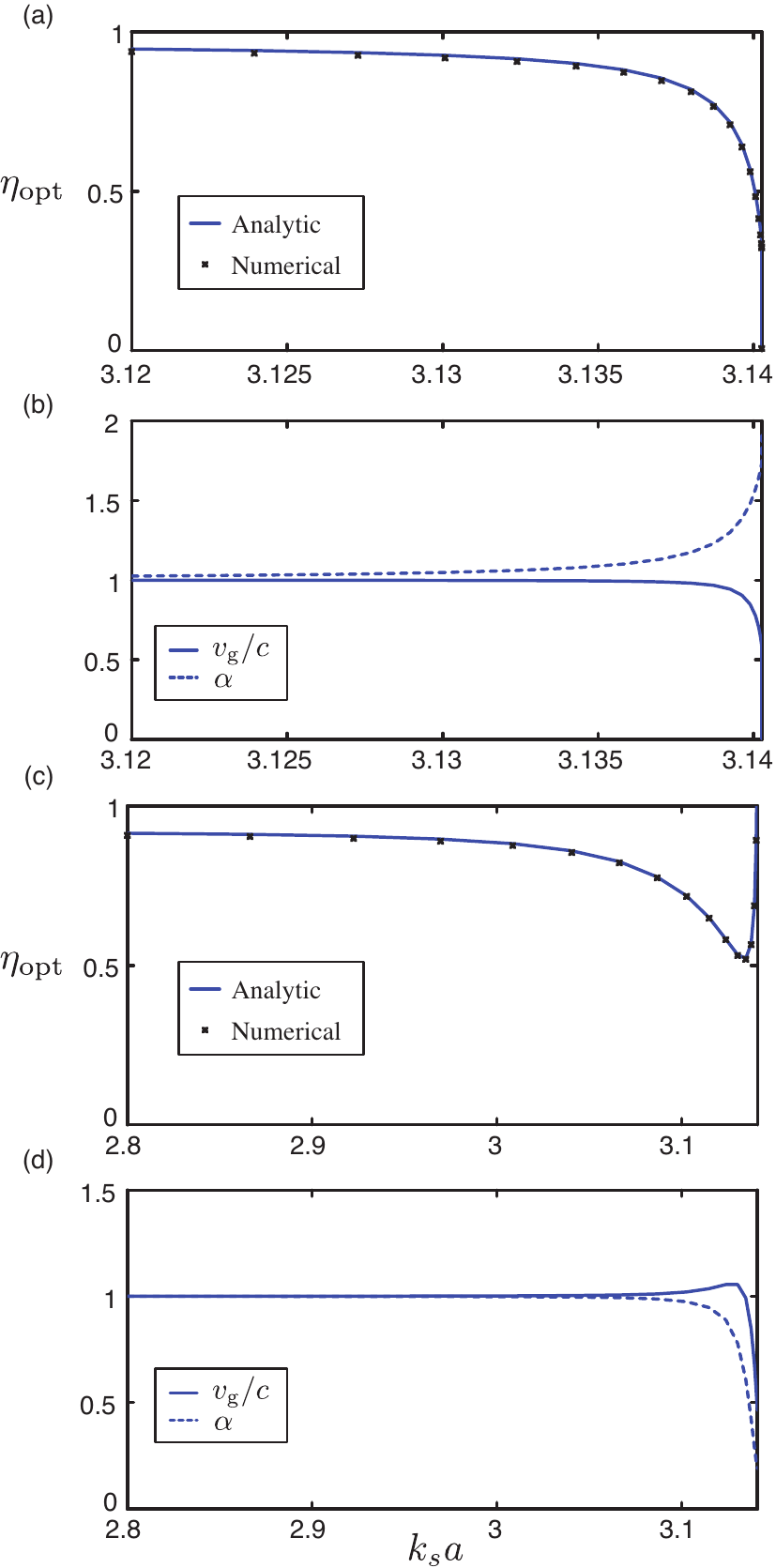}
\caption{(Color online) Storage efficiency, group velocity and overlap near a band gap. (a) The storage efficiency of the Raman memory is plotted as a function of the signal frequency near the band gap. The blue solid line is found from the analytic kernel in Eq.~(\ref{Green_give}); the black crosses are found from a numerical solution of Eqs.~(\ref{motionA}), (\ref{motion}), which takes account of walk-off between the signal and control. The efficiency falls as the signal frequency approaches the band edge. (b) The corresponding variation of both the group velocity, which falls, and the overlap parameter $\alpha$, which increases as the band edge is approached. (c) EIT storage efficiency, which falls, and then rises sharply as the signal field approaches the band edge. In part (d) we plot the group velocity, which rises and then falls, and the overlap parameter $\alpha$, which falls, as the band edge is approached.}
\label{fig:EIT_Raman}
\end{center}
\end{figure}

Figure~\ref{fig:EIT_Raman} shows the variation of the optimal storage efficiency $\eta_\mathrm{opt}$, along with the group velocity $v_\mathrm{g}$ and the overlap parameter $\alpha$, for the two memory protocols, as the frequency of the signal field approaches a band edge. In the Raman memory, the group velocity falls steadily to zero, and the overlap increases, consistent with the transition via Bragg scattering of the signal carrier from a plane to a standing wave with nodes between the lattice sites. Both of these effects might be expected to increase the memory efficiency, but instead the memory efficiency actually \emph{falls}.

The reason for this surprising negative result appears to be related to the joint atomic/optical character of the memory interaction: although close to a band edge the coupling of the signal field to the atomic polarization increases, the coupling of the signal field to the storage state $\ket{3}$ is unchanged, since this is mediated by the control field, which experiences no dispersion. The atomic polarization simply re-radiates the signal field at an enhanced rate, and less of the signal is mapped into the desired storage state.

The behaviour of the EIT memory is quite different. Here, the modulation of the potential $V(z)$ is purely imaginary, since $\Delta=0$. This gives the band structure a radically different appearance (see part (b) of Figure~\ref{fig:dispersion}), and this is manifested in the variation of the group velocity, which becomes superluminal briefly, before falling sharply, as shown in part (d) of Figure~\ref{fig:EIT_Raman}. The overlap parameter also falls, in contrast to the Raman case, suggesting that at the band edge the EIT carrier becomes a standing wave with nodes aligned with the lattice sites. The optimal efficiency changes rather counter-intuitively. It first decreases, and then rises sharply. We understand this behaviour in the following way.

\begin{figure}[h]
\begin{center}
\includegraphics[width=\columnwidth]{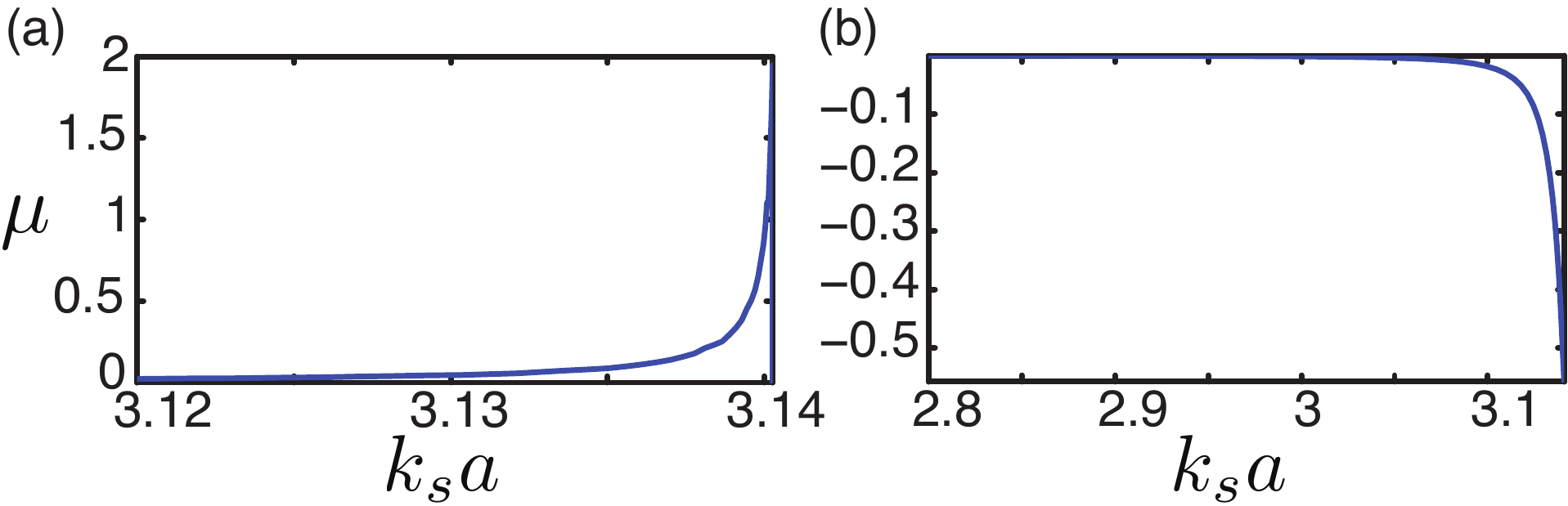}
\caption{(Color online) Damping. (a) The variation of the damping parameter $\mu = 1-\mathrm{Im}\left\{k\right\}L/\mathrm{Re}\left\{d\gamma/\Gamma \right\}$ for the Raman memory, as a function of the signal frequency, approaching the bandgap. In the absence of band structure, $\mu=0$, but the increase in $\mu$ close to the band edge is associated with increased absorption of the signal carrier wave. (b) The corresponding variation of $\mu$ for the EIT memory. Close to the band edge, $\mu$ becomes increasingly negative, which is associated with reduced absorption. This phenomenon is known as anomalous transmission \cite{Oberthaler:1996fk,Batelaan:1997uq,Berry:1998kx}.}
\label{fig:damping}
\end{center}
\end{figure}

In an EIT memory, the coupling of the signal to the atomic polarization is directly related to the memory efficiency; the control field dresses the atoms and mixes the storage and excited states, excitation of the atomic polarization then directly excites the storage state. Therefore the initial reduction in $\alpha$ and the increase in $v_\mathrm{g}$ act to decrease the memory efficiency. The subsequent `turnaround' in efficiency very near to the band edge is partly explained by the sharp reduction in $v_\mathrm{g}$, which serves to increase the atom-signal coupling. But a second effect is more significant: \emph{anomalous transmission}. This refers to a reduction in the absorption that is characteristic of an imaginary potential, when close to a forbidden band --- it has been observed in the transmission of atoms through an optical standing wave \cite{Oberthaler:1996fk,Batelaan:1997uq,Berry:1998kx}; here it pertains to the transmission of an optical field through a periodic array of atoms. The attenuation of the signal carrier for the Raman and EIT protocols is shown in Figure~\ref{fig:damping}. In the absence of any modulation, $m(z) = 1$, and Eq.~(\ref{eig}) can be solved trivially, yielding $\mathrm{Im}\left\{k\right\} = \mathrm{Re}\left\{d\gamma/\Gamma L\right\}$. Any departure from this value is associated with the band structure arising from the lattice modulation. We quantify these departures by defining the dimensionless damping parameter $\mu = 1-\mathrm{Im}\left\{k\right\}L/\mathrm{Re}\left\{d\gamma/\Gamma\right\}$. The variation of $\mu$ as the signal frequency approaches a band edge is plotted in Figure~\ref{fig:damping} for both memory protocols. In the Raman case, the damping increases near the forbidden band, consistent with the transition of the signal carrier from a propagating to an evanescent wave. In the case of EIT however, the damping \emph{drops}, becoming negative near the band gap. The absorptive scattering of the signal is thus dramatically reduced, and it is this anomalous transmission \cite{Oberthaler:1996fk,Batelaan:1997uq,Berry:1998kx} that explains the sudden increase in the optimal storage efficiency when very near the band edge.

Here we note that our theoretical model has limited applicability so close to the band edge. The slowly varying envelope approximation used in deriving Eqs.~(\ref{Aphi}), (\ref{Bphi}) requires that a single carrier wave correctly describes the propagation of all the frequencies comprising the signal pulse. When the dynamics become extremely dispersive, as they do in very close proximity to the stop band, the bandwidth of the signal pulse will itself span a range of carrier modes. The detailed analysis of this effect lies beyond the scope of the present treatment, but in any case it will lead to the dispersive break-up of the signal pulse, and to a reduced memory efficiency. Therefore one should probably not conclude that enhanced storage efficiency can be achieved via EIT by tuning very close to a band edge. In the next section we show that reflection losses negate any such enhancement in any case.

\section{Reflection}
A final important consideration for the lattice memory is the reflectivity of the interface between free space and the ensemble \cite{Kozlov:2002vn,Petrosyan:2007vn,Fleischhauer:2002ph}. Although the optimal efficiency of EIT appears to approach unity very near the band edge, it seems that the portion of the signal field lost due to reflection at the entrance face of the memory is sufficiently large to abrogate any advantage.

\begin{figure}[h]
\begin{center}
\includegraphics[width=\columnwidth]{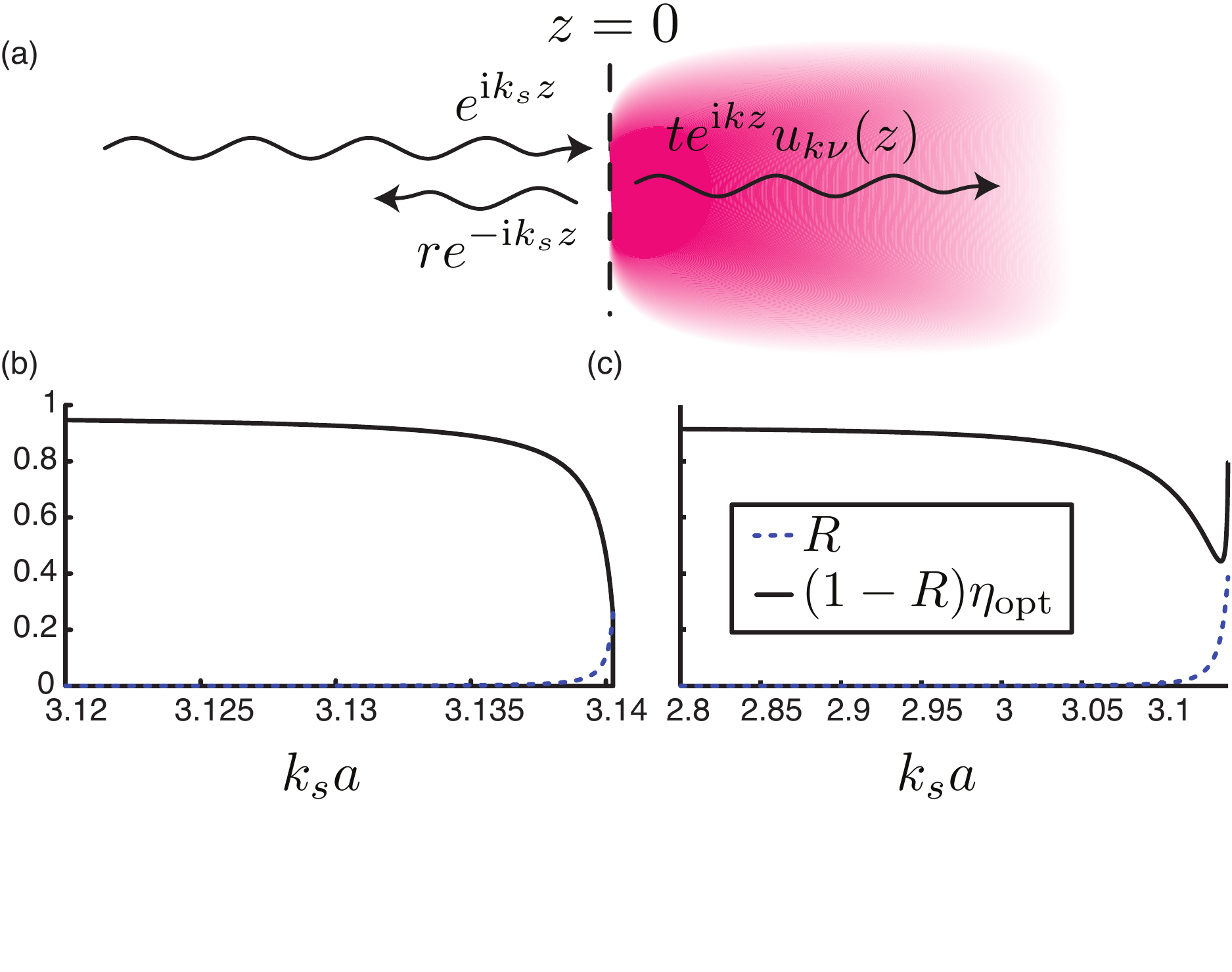}
\caption{(Color online) Reflection from the entrance face. (a) continuity of the incident, reflected and transmitted signal carrier waves at the entrance face of the lattice determine the reflectivity $R = |r|^2$ of the memory. (b) the variation of $R$, and the corresponding storage efficiency $(1-R)\eta_\mathrm{opt}$, as the signal frequency approaches the band edge, for the Raman memory. (c) the reflectivity and resulting efficiency of the EIT protocol. Both memory protocols suffer a dramatic reduction in efficiency close to the band edge, due to a sharp rise in the reflectivity of the lattice.}
\label{fig:reflect}
\end{center}
\end{figure}

We obtain an expression for the reflectivity of the ensemble by imposing continuity of the signal carrier waves and their derivatives at the free-space/lattice interface (see part (a) of Figure~\ref{fig:reflect}). The result is
$$
R = \left|\frac{r_1-r_2}{r_1+r_2}\right|^2,
$$
where $r_1 = k_su_{k\nu}(0)$, and $r_2 = ku_{k\nu}(0)-\mi \partial_z u_{k\nu}(z)|_{z=0}$. Parts (b) and (c) of Figure~\ref{fig:reflect} show the dependence of the reflectivity, along with the appropriately attenuated storage efficiency $(1-R)\eta_\mathrm{opt}$, on the proximity of the signal frequency to the band edge for the two memories.
The reflectivity $R$ increases dramatically as the signal frequency approaches the band gap, for both the EIT and Raman memory protocols. The fraction of the signal that penetrates the ensemble and is subsequently stored is therefore reduced, and the overall memory efficiency suffers at the band edge. In the case of the Raman memory, this effect compounds the reduction in efficiency arising from the dispersive propagation. For the EIT memory, the sharp rise in efficiency due to anomalous transmission is tempered by the reflection losses, so that there is no longer any advantage in tuning the signal close to a band edge.

\section{Conclusion}
We have analyzed the effects of band structure on the efficiency of ensemble-based memory protocols in optical lattices. Although a number of interesting effects, including enhanced atom-light coupling, sub and superluminal group velocities and anomalous transmission emerge from the model, the memory efficiencies are nonetheless reduced near a band edge. On the basis of this analysis, we conclude that one should avoid tuning the optical fields too close to a band edge when performing light storage in an optical lattice. Since a deep lattice potential is achieved by tuning the trapping lasers close to resonance, there is a real possibility of `accidental' coincidence of the band edges with the signal frequency; this should be avoided if possible.

 Far from the forbidden bands, the memory efficiency in a lattice is the same as would be achieved in an equivalent disordered ensemble with the same optical depth. Of course, the long atomic coherence times available in optical lattices make them one of the most promising technological routes to useful quantum optical memories.

\acknowledgements
This work was supported by the EPSRC through the QIP IRC
(GR/S82716/01) and project EP/C51933/01. JN
thanks Hewlett-Packard. IAW was
supported in part by the European Commission under the Integrated
Project Qubit Applications (QAP) funded by the IST directorate as
Contract Number 015848, and the Royal Society.

%\bibliography{../../../references/references}

\end{document}